\documentclass[a4paper,fleqn,usenatbib]{mnras}
\usepackage{natbib}

\usepackage{newtxtext,newtxmath}

\usepackage[T1]{fontenc}
\usepackage{ae,aecompl}

\usepackage{graphicx}	
\usepackage{amsmath}	
\usepackage{amssymb}	
\usepackage{booktabs}
\usepackage{color}

\usepackage [english]{babel}
\usepackage [autostyle, english = american]{csquotes}
\MakeOuterQuote{"}

\title[Vortex and Flux Tube Arrays in a Neutron Star]{Stability of interlinked neutron vortex and proton flux tube arrays in a neutron star. II. Far-from-equilibrium dynamics}

\author[L. V. Drummond and A. Melatos]{
L. V. Drummond\thanks{E-mail: l.drummond@student.unimelb.edu.au}
and A. Melatos\thanks{E-mail: amelatos@unimelb.edu.au}
\\
School of Physics, University of Melbourne, Parkville, VIC 3010, Australia\\
Australian Research Council Centre of Excellence for Gravitational Wave Discovery (OzGrav)\\
}

\date{Accepted XXX. Received YYY; in original form ZZZ}

\pubyear{2017}

\begin{document}

\label{firstpage}
\pagerange{\pageref{firstpage}--\pageref{lastpage}}
\maketitle


\begin{abstract}
The equilibrium configurations of neutron superfluid vortices interacting with proton superconductor flux tubes in a rotating, harmonic trap are non-trivial in general, when the magnetorotational symmetry is broken. A non-zero angle $\theta$ between the magnetic and rotation axes leads to tangled vorticity due to competition between vortex-vortex repulsion and vortex-flux-tube pinning. Here we investigate the far-from-equilibrium behaviour of the vortices, as the trap decelerates, by solving the time-dependent, stochastic, Gross-Pitaevskii equation numerically in three dimensions. The numerical simulations reveal new vortex behaviours. Key geometrical attributes of the evolving vortex tangle are characterised, as is the degree to which pinning impedes the deceleration of the neutron condensate as a function of $\eta$, the pinning strength, and $\theta$. The simulated system is a partial analogue of the outer core of a decelerating neutron star, albeit in a very different parameter regime.
 \end{abstract}




\section{Introduction}


The outer core of a neutron star, defined as the layer where the mass density $\rho$ spans the range $1.6\lesssim\rho/(10^{14}\mbox{g}\ \mbox{cm}^{-3})\lesssim3.9$, contains a condensed mixture of superfluid neutrons ($\gtrsim 95\%$ by mass), superconducting protons and viscous electrons \citep{Yakovlev1999,LivingReview}. The proton and electron components are magnetically locked to, and hence co-rotate with, the rigid crust. Quantised vortices cause the neutron superfluid to rotate. Each vortex carries a quantum of circulation, $\kappa= h/(2m_n)=1.98 \times 10^{-3} \  \mbox{cm}^2 \ \mbox{s}^{-1}$, and there are $N_v=10^{16}(\Omega/ \mbox{rad} \ \mbox{s}^{-1})$ vortices in a typical neutron star, where $\Omega$ is the angular velocity \citep{LivingReview}. The proton superconductor is typically considered to be type-II, meaning that magnetic flux penetrates its volume in the form of flux tubes. Each flux tube carries a magnetic flux quantum $\Phi_0= hc/(2e)=2.07 \times 10^{-7}\ \mbox{G} \ \mbox{cm}^{2}$ \citep{Mendell1991,ChauDing}, and there are $N_{\Phi}=10^{30}(B/10^{12} \mbox{G})$ flux tubes in a typical neutron star, where $B$ is the magnetic field strength. Neutron vortices and proton flux tubes repel strongly according to many theoretical calculations through density and magnetic interactions caused by neutron-proton entrainment \citep{Andreev1975,Glamp2011}, with an interaction energy of $\sim \mbox{MeV}$ per junction \citep{Sauls1989,Srivasan1990,Bhatta1991,Ruderman1998}. It is therefore natural to assume that the vortices and flux tubes cannot pass through each other easily, if the star's rotation and magnetic axes are misaligned \citep{Ruderman1998}. 

The presumption that neutron vortices and proton flux tubes are locked together globally relies on the vortices and flux tubes remaining rectilinear on macroscopic, stellar scales, like two inclined grids of stiff wires. However, it is known that the neutron vortex array is disrupted into a tangle under a range of plausible conditions. In a previous paper, we computed the \textit{equilibrium} configurations of neutron vortices interacting with a static flux tube array \citep{DrummondMelatos2017}. We found that a non-zero angle $\theta$ between the global magnetic and rotation axes ($\boldsymbol{\Omega}$ and $\mathbf{B}$ respectively) leads to tangled vorticity. The increases in line length and curvature, when vortices become tangled for $\theta\neq0^{\circ}$, are characteristic of instability in models of superfluid turbulence \citep{Barenghi2001}. Superfluid turbulence produces polarised vortex tangles \citep{SFNSturbulence}, as does an instability analogous to grid turbulence, which occurs when vortices are pushed against the rigid flux tube array \citep{Link2012}. 

There is theoretical evidence to support the premise that the fluid in the interior of a neutron star exhibits superfluid turbulence in two forms: macroscopic, Kolmogorov-like eddies \citep{MelatosPeralta2010} and microscopic vortex tangles, where instabilities drive growth of helical Kelvin waves, leading to vortex reconnection on microscopic scales \citep{Peralta2006,AnderssonSidery2007}. In this paper, we are primarily concerned with the latter phenomenon. We assess its implications in a neutron star context by investigating the far-from-equilibrium behaviour in response to stellar braking. A three-dimensional investigation of neutron star spin down for misaligned $\boldsymbol{\Omega}$ and $\mathbf{B}$ has the capacity to reveal new vortex behaviours. As the stellar crust spins down electromagnetically, vortices move radially outwards until they reach the edge of the condensate and annihilate at the boundary. We aim to study the vortex motion as the star spins down. Do the flux tubes impede the outward vortex drift, as commonly assumed, or can the vortices slip past more easily because they are tangled and segmented? Does a vortex unpin from a flux tube in small segments rather than all at once \citep{LinkEpstein1991}? 

The paper is structured as follows. Section \ref{section:GPmodel} presents an overview of the model describing the evolution of the neutrons according to the Gross-Pitaevskii equation (GPE) driven by a spin-down torque. In Section \ref{section:vortexcrystal}, the properties of the initial vortex configuration ("vortex crystal") are examined, before spin down is switched on. In Section \ref{section:vortexevo}, the far-from-equilibrium dynamics of the system during spin down are computed, and features of the evolving vortex structure are discussed. In Section \ref{section:tangle}, the geometry of the vortex tangle is characterised. In Section \ref{section:rotation}, we calculate how strongly vortex-flux-tube interactions affect the observable deceleration of the crust as a function of $\theta$ and the pinning potential. Section \ref{section:limitations} outlines the limitations of the model. We emphasize that the dynamic range captured in the simulations is not realistic for a neutron star, even though we ensure that the ordering of relevant dimensionless ratios is consistent astrophysically. As in \citet{DrummondMelatos2017}, we omit out of computational necessity a vital piece of physics: the self-consistent response of the flux tubes. Including the latter effect is a challenge for future numerical work.

\section{Gross-Pitaevskii Model of the Outer Core}
    \label{section:GPmodel}
We describe the neutron condensate in the outer core using the Gross-Pitaevskii equation (GPE). The GPE has been employed previously to successfully simulate the triggering of vortex avalanches to produce rotational glitches, achieving avalanche statistics in line with observational data \citep{LilaMelatos2011, LilaMelatos2012,Douglass2015}. In addition, the GPE has been used to simulate knock-on mechanisms that govern the avalanches \citep{LilaMelatos2013}. Despite the above successes, we stress that the GPE model is idealised in a number of ways, which are itemized in detail by \citet{DrummondMelatos2017}. (i) The neutrons are not dilute; the s-wave scattering length $a_s$ and the average neutron separation are of the same order of magnitude. (ii) The GPE describes a weakly interacting Bose gas, while neutron matter is in the strongly interacting regime. (iii) The viscous electron plasma in the outer core is not treated in our model \citep{Mendell1998, Glamp2011}. (iv) Pinning in the inner crust is omitted \citep{Donati2006,Avogadro2007,Avogadro2008,Grill2012,Haskell2013}. The reader is referred to Section \ref{section:limitations} for further details about the limitations of the model. 

In this paper, we solve the "stochastic GPE" \citep{GardinerStochastic}
\begin{equation} 
\label{eq:realGP} 
(\mathrm{i}-\gamma)\frac{\partial\psi}{\partial t}	= \left(-\frac{1}{2}\nabla^{2}+V+|\psi|^{2}-\Omega \hat{L}_{z}+\mathrm{i}\gamma \mu \right)\psi+\mathcal{H}_{int}[\psi,\phi]
\end{equation} 
 in dimensionless form, where $\psi$ is the neutron order parameter, and $\phi$ is the proton order parameter. $\psi$ is normalised to give $\int|\psi|^2\mbox{d}^3\mathbf{x}=N_n$ and $\phi$ to give $\int|\phi|^2\mbox{d}^3\mathbf{x}=N_p=0.05N_n$ \citep{LivingReview}, where $N_n$ and $N_p$ are the total numbers of condensed neutrons and protons respectively. The dimensionless independent variables in equation (\ref{eq:realGP}) are normalized with respect to length- and time-scales $\xi_n=\hbar/(2m_nn_{n}U_{0})^{1/2}$ and $\tau=\xi_n/c_{s}=\hbar/(n_{n}U_{0})$, where $U_{0} $ is the strength of the neutron self-interaction, $n_n$ is the background neutron number density, $\xi_n$ is the neutron coherence length and $c_s$ is the speed of sound. $\psi$ and $\phi$ are expressed in units of $n_n^{1/2}$. The term $\mathrm{i}\gamma\mu\psi$ in (\ref{eq:realGP}) introduces a phenomenological coupling to a thermal reservoir, where $\gamma$ is a damping constant, and $\mu$ is the chemical potential. The physics and regime of validity of the stochastic GPE are discussed in detail elsewhere \citep{GardinerStochastic, Douglass2015, DrummondMelatos2017}. Equation (\ref{eq:realGP}) is written in the frame co-rotating with the crust at angular velocity $\Omega$, $\hat{L}_{z}$ is the angular momentum operator, $V$ is a cylindrically symmetric trapping potential, given by $V=\tilde{\omega}^{2}(x^{2}+y^{2})/2$, and $\mathcal{H}_{int}[\psi,\phi]$ describes the density and current-current interactions between the neutron superfluid and the proton superconductor [see equation (\ref{eq:Hint}) below].


We describe the superconducting protons with a static ansatz resembling numerical solutions of the Ginzburg-Landau equations \citep{Clem,Tinkham}. We model the flux tube array as the product of single-flux-tube wavefunctions $\phi(\mathbf{x})=\prod_{i}\phi_f(\mathbf{x}-\mathbf{x}_i)$ and the linear superposition of single-flux-tube vector potentials $\mathbf{A}(\mathbf{x})=\sum_{i}\mathbf{A}_f(\mathbf{x}-\mathbf{x}_i)$, where $\mathbf{x}_i$ is the position of the $i^{\mbox{th}}$ flux tube in the mid-plane of the system. For an isolated flux tube at the origin, we have \citep{Clem}
 \begin{equation}
  \label{eq:phi} 
 \phi_f=n_p^{1/2}(r/\tilde{r})^2\mbox{e}^{\mathrm{i}\chi},
  \end{equation}
  
   \begin{equation}
     \label{eq:Af} 
 \mathbf{A}_f 	=	\frac{\Phi_{0}}{2\pi r} \left[1-\frac{\tilde{r}K_{1}\left(\tilde{r}/\lambda\right)}{\sqrt{2}\xi_p K_{1}\left(\sqrt{2}\xi_p/\lambda \right)} \right] \hat{\chi},
  \end{equation}
 in dimensional form, where $(r,\chi)$ are polar co-ordinates, $n_p$ is the background proton number density, $\xi_p$ is the proton coherence length, $\lambda$ is the London penetration depth, we write $\tilde{r}=(r^2+2\xi_p^2)^{1/2}$, and $K_{n}(x)$ is a modified Bessel function of the second kind of order $n$. Equations (\ref{eq:phi}) and (\ref{eq:Af}) are a good approximation, when flux tubes are separated by more than $d_{\Phi}\approx 5\xi_p$, where $d_{\Phi}$ is the distance between individual flux tubes \citep{BrandtRev}.
 
We consider two proton-neutron interaction mechanisms for simplicity. \citet{Sauls1989} identified a pinning interaction between flux tubes and vortices that arises from the difference in condensation energy between the pinned and free configurations, due to local modifications in the condensation energy associated with the density perturbation in the core of the flux tube. Secondly, the many-body effect called "entrainment" causes neutrons and protons to drag along clouds of quasi-particles of the opposite species \citep{Andreev1975,Sjoberg1976,Andersson2007,Glamp2011}. We model these effects in (\ref{eq:realGP}) via terms of the form

\begin{equation}
\label{eq:Hint} \mathcal{H}_{int}[\psi,\phi]=\eta|\phi|^{2}\psi - \frac{\mathrm{i}\zeta}{2}\left(2\mathbf{j}_p\cdot\nabla\psi+\psi\nabla\cdot\mathbf{j}_p\right), 
\end{equation}
with
 \begin{equation}
 \label{eq:nodimjp}
\mathbf{j}_p=\frac{\mathrm{i}}{2}\left[\phi\left(\nabla+\mathrm{i}\xi_{n}\frac{2e}{\hbar c}\mathbf{A}\right)\phi^{*}-\phi^{*}\left(\nabla-\mathrm{i}\xi_{n}\frac{2e}{\hbar c}\mathbf{A}\right)\phi \right]
 \end{equation}
where $\eta$ and $\zeta$ are the dimensionless density and current-current coupling coefficients respectively. The reader is referred to \citet{DrummondMelatos2017} for more information. In this work, we do not solve the full dynamical problem with respect to the protons; entrainment of the protons around the neutron vortices is omitted, to be addressed in future work. Therefore, an important open question remains as to whether vortices or flux tubes "lead" the interaction. We discuss this further in Section \ref{section:limitations}.

The equilibrium state of the system is calculated by solving (\ref{eq:realGP}) in imaginary time ($t\,\to\,-\mathrm{i}t$, $\gamma\to0$) \citep{DrummondMelatos2017}. When $\mathbf{B}$ and $\boldsymbol{\Omega}$ are misaligned, the system is frustrated by the competition between vortex-vortex repulsion and vortex-flux-tube pinning, resulting in the development of vortex tangles \citep{DrummondMelatos2017}. \citet{DrummondMelatos2017} found that states with misaligned $\mathbf{B}$ and $\boldsymbol{\Omega}$ are frustrated, leading to a protracted and "glassy" evolution in imaginary time before the absolute minimum energy is achieved. The convergence criterion $\int \mbox{d}^3 \mathbf{x} \ |\Delta \psi|^2 \leq10^{-5} N_n$, where $N_n$ is the number of neutrons in the simulation box, quantifies the system's convergence to the minimum energy state, where $\Delta \psi$ is the difference in the wavefunction between successive imaginary time steps. 

We use the equilibrium state as the starting point for the time-dependent solutions of (\ref{eq:realGP}). We spin down the system to simulate the electromagnetic braking a neutron star experiences during its lifetime \citep{LilaMelatos2011,LilaMelatos2012,Douglass2015}. The back-reaction from the condensate is included along with the braking torque, viz.
\begin{equation}
\label{eq:brake}
I_c\frac{\mbox{d}\Omega}{\mbox{d}t}=-\frac{\mbox{d}\langle\hat{L}_z\rangle}{\mbox{d}t}+N_{\mathrm{em}},
\end{equation}
where $N_{\mathrm{em}}<0$ is the electromagnetic braking torque, $\mbox{d}\langle \hat{L}_z \rangle/\mbox{d}t$ is the rate of change of the condensate's angular momentum and $I_c$ is the moment of inertia of the rigid crust. We solve (\ref{eq:realGP}) using a spectral Runge-Kutta method in the interaction picture (RK4-IP). The algorithm is well-suited to solving time-dependent non-linear Schrödinger equations like the GPE \citep{CaradocDaviesThesis,Balac2013}. As in \citet{DrummondMelatos2017}, we select $\int \mbox{d}^3 \mathbf{x} \ |\Delta \psi|^2 \leq10^{-5} N_n $ as the (arbitrary) numerical tolerance where convergence is achieved in obtaining the initial states for the simulations in this paper. We refer the reader to \citet{DrummondMelatos2017} for further details about the convergence metric $\int \mbox{d}^3 \mathbf{x} \ |\Delta \psi|^2$.

\section{Initial State: Vortex Crystal}
\label{section:vortexcrystal}

\begin{figure}
\centering
\hspace*{-3em}
\includegraphics[height=44em]
{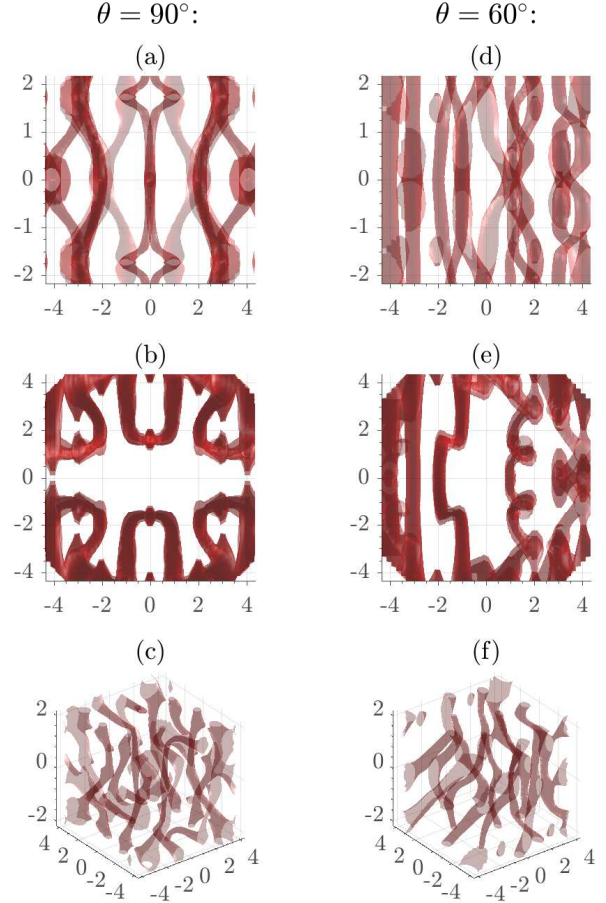}
\caption[Protons]
{Different projected views of the vortex crystal produced by imaginary time evolution at $t=0$ for $\theta=90^{\circ}$ (left column) and $\theta=60^{\circ}$ (right column). (a) and (d). Side-on view, line of sight perpendicular to $\mathbf{\Omega}$. (b) and (e). Bird's eye view, line of sight parallel to $\mathbf{\Omega}$. (c) and (f). Line of sight elevated by $30^{\circ}$ from the $x\mbox{-}y$ plane. The red shading signifies the vortices, drawn where $|\psi|^2$ drops below 10\% of its maximum. Length units are $\xi_n$. Parameters: $\Omega(t=0)=0.5$, $\eta=-100$, $d_{\Phi}=2$, $\gamma=0.001$, $N_{\mathrm{em}}=-0.006I_c$, $\tilde{N}_n=N_n/(n_n\xi_n^{3})=8\times10^3$ and $\tilde{N}_p=0.05\tilde{N}_n$.}
\label{fig:vortexcrystal}
\end{figure}

 \begin{figure}
\centering
\hspace*{-3em}
\includegraphics[height=27em]
{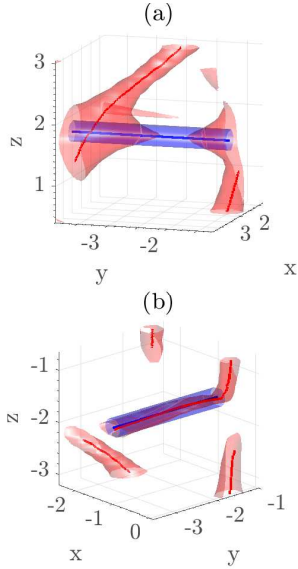}
\caption[Protons]
{Deformation and pinning of vortices in the equilibrium vortex crystal. Close-up snapshot of (a) vortex "spike" and (b) a pinned vortex at $t=0$ for $\theta=90^{\circ}$. The red (blue) shading signifies the vortices (flux tubes), drawn where $|\psi|^2$ ($|\phi|^2$) drops below 10\% of its maximum. The solid red (blue) lines denote the cores of the vortices (flux tubes). Length units are $\xi_n$. Parameters: $\Omega(t=0)=0.5$, $\eta=-100$, $d_{\Phi}=2$, $\gamma=0.001$, $N_{\mathrm{em}}=-0.006I_c$, $\tilde{N}_n=N_n/(n_n\xi_n^{3})=8\times10^3$ and $\tilde{N}_p=0.05\tilde{N}_n$.}
\label{fig:spike}
\end{figure}

 \begin{figure}
\centering
\hspace*{-3em}
\includegraphics[height=13em]
{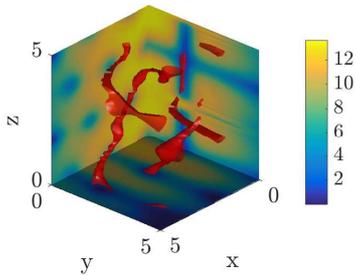}
\caption[Protons]
{Vortex configuration and $|\psi|^2$ contours in the equilibrium ground state of the vortex crystal. Close-up snapshot of the neutron vortices (as depicted in Figure \ref{fig:vortexcrystal}) and $|\psi|^2$ contours (the blue-, green- and yellow-shaded walls of the box) at $t=0$ for $\theta=90^{\circ}$. The red shading signifies the vortex lines, drawn where $|\psi|^2$ drops below 10\% of its maximum. The walls of the box show $|\psi|^2$ cross-sections (in units of $n_n$) for the midplanes $x=0$, $y=0$ and $z=0$. Length units are $\xi_n$. A colour bar is provided for the contour values in each panel in units of $n_n$. Parameters: $\Omega(t=0)=0.5$, $\eta=-100$, $d_{\Phi}=2$, $\gamma=0.001$, $N_{\mathrm{em}}=-0.006I_c$, $\tilde{N}_n=N_n/(n_n\xi_n^{3})=8\times10^3$ and $\tilde{N}_p=0.05\tilde{N}_n$.}
\label{fig:backgrounddensity}
\end{figure}

The initial vortex configuration for $\theta\neq0^{\circ}$ which emerges following imaginary time evolution has some unusual properties. It represents a compromise  between the propensity for global vortex alignment along $\mathbf{\Omega}$ and the local pinning forces along $\boldsymbol{B}$. The vortices do not crystallise into a rectilinear array as they do for $\theta=0^{\circ}$. Instead, they settle into a complex, partially polarised, interlocked pattern. Examples are shown in Figure \ref{fig:vortexcrystal}, where isosurfaces of $|\psi|^2$ are plotted, for $\theta=90^{\circ}$ and $\theta=60^{\circ}$.

This behaviour is similar to that of nuclear pasta in neutron star crusts, which also have complex geometries due to competing interactions \citep{Ravenhall1983,Hashimoto1984}. Nuclear pasta phases are expected at the base of the inner crust, where the density is high enough that individual atomic nuclei begin to touch. Here, the competition between nuclear attraction and Coulomb repulsion reorganises neutrons and protons into complex patterns such as sheets (lasagna) or tubes (spaghetti) \citep{Caplan2016}. The interplay between competing lattices has been investigated in two dimensions by superimposing an optical lattice on a rotating BEC, leading to competition between the triangular lattice arising from the mutual repulsion of the vortices and the pinning forces due to the optical square lattice \citep{Tung2006}.

An example of a canonical vortex tangle is the irregular, isotropic arrangement of vortices  generated by evolving the GPE starting from uniform initial density and  spatially random initial phase \citep{Tsubota2009}. The $\theta=90^{\circ}$ state shares similarities with a canonical vortex tangle, in the sense that the vortices are quasi-isotropic. Viewed with the line of sight elevated $30^{\circ}$ above the $x\mbox{-}y$ plane, e.g. Figure \ref{fig:vortexcrystal}(c), the structure appears isotropic. However, viewing it along the rotation axis [Figure \ref{fig:vortexcrystal}(b)] or along the magnetic axis [Figure \ref{fig:vortexcrystal}(a)] reveals latent symmetries. It is more accurate to characterise this state as a kind of "vortex crystal" with two planes of symmetry. Moving from the special case $\theta=90^{\circ}$ to $\theta=60^{\circ}$ yields a more irregular vortex arrangement (see Figure \ref{fig:vortexcrystal}). In Section \ref{section:tangle}, we introduce the tangle polarity $\mathbf{p}$, defined in (\ref{eq:polarity}), which measures the isotropy of the tangle; $|\mathbf{p}|=1$ indicates perfect polarisation and $|\mathbf{p}|=0$ indicates perfect isotropy. The vortex crystal for $\theta=90^{\circ}$ has $|\mathbf{p}|=0.625$ (see Figure \ref{fig:vortexproperties}). When the system is spun down in Section \ref{section:vortexevo},  $|\mathbf{p}|$ reduces to 0.175 by the end of the simulation ($\theta=90^{\circ}$). We discuss the evolution of $\mathbf{p}$ in detail in Section \ref{section:tangle}.

The flux tubes run along the magnetic field $\mathbf{B}$ as unbroken troughs. Perfect vortex-flux-tube pinning cannot be achieved, because the vortices must align globally with $\boldsymbol{\Omega}$. Both the internal vortex structure and the global geometry are deformed by the presence of the flux tube array, as illustrated in Figure \ref{fig:spike}. Low-$|\psi|^2$ ($0\leq|\psi|^2 \leq 4n_n$) "spikes" stick out from the vortices, where vortices intersect the flux tube array. Equation (\ref{eq:Hint}) favours the overlap of low-$|\psi|^2$ regions with low-$|\phi|^2$ regions, so, rather than pinning to a flux tube, a vortex may grow a spike at a vortex-flux-tube junction.  A close-up of a spike is presented in Figure \ref{fig:spike}(a). In addition, portions of the vortices align along the flux tube array, as show in Figure \ref{fig:spike}(b). In Figure \ref{fig:spike}(a), where a spike is created, the vortex core (denoted by a solid red line) does not run along the flux tube. In Figure \ref{fig:spike}(b), the vortex core does run along the flux tube, signifying vortex-flux-tube pinning. 

\begin{figure*}
\centering
\includegraphics[height=30em]
{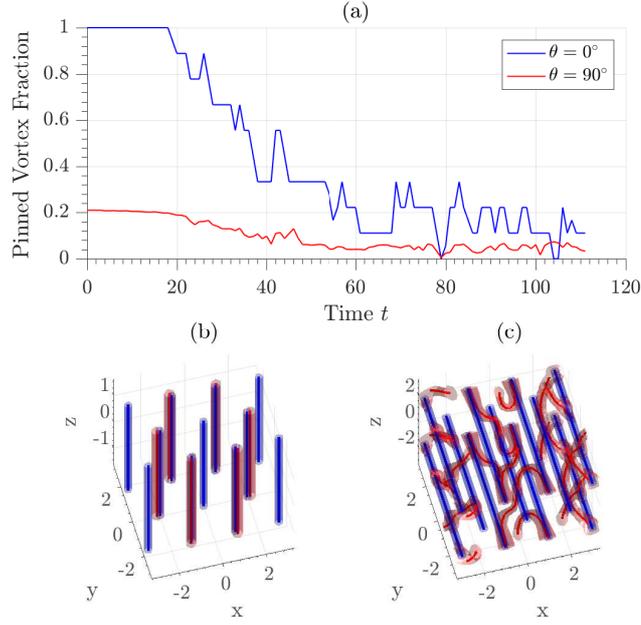}
\caption[Protons]
{Vortex pinning to the flux tube array. (a) The fraction of vortex points that are pinned versus time for $\theta=0^{\circ}$ (blue curve) and $\theta=90^{\circ}$ (red curve). (b) Vortex and flux tube structure at $t=0$ for $\theta=0^{\circ}$. (c) Vortex and flux tube structure at $t=0$ for $\theta=90^{\circ}$. The red (blue) shading signifies the vortices (flux tubes), drawn where $|\psi|^2$ ($|\phi|^2$) drops below 10\% of its maximum. The solid red (blue) lines denote the cores of the vortices (flux tubes). Length units are $\xi_n$. Parameters: $\Omega(t=0)=0.5$, $\eta=-100$, $d_{\Phi}=2$, $\gamma=0.001$, $N_{\mathrm{em}}=-0.006I_c$, $\tilde{N}_n=N_n/(n_n\xi_n^{3})=8\times10^3$ and $\tilde{N}_p=0.05\tilde{N}_n$. In (b) and (c), we plot a close-up of the system, containing the six central vortices only, to avoid overcrowding the plot.}
\label{fig:pinnedfraction}
\end{figure*}

We quantify the extent to which vortices are pinned to flux tubes by identifying contiguous subsets of points along the vortex, as described in \citet{DrummondMelatos2017}. For every point along each vortex, we conduct a fixed-radius search\footnote{We use the MATLAB function \texttt{rangesearch} to return all neighbours within a specified distance.}. We classify a vortex point as pinned, if it lies within $0.4\xi_n$ of a flux tube point. We find that, at $t=0$, the fraction of pinned vortex points is 1.0 (0.210) for $\theta=0^{\circ}$ ($\theta=90^{\circ}$). Figure \ref{fig:pinnedfraction} in Section \ref{section:vortexevo} shows how the pinned vortex fraction changes during the simulation.

We examine the density distribution of the vortex crystal in detail in Figure \ref{fig:backgrounddensity}. The figure presents a close-up of the isosurfaces in Figure \ref{fig:vortexcrystal}. The walls of the box show cross-sections of $|\psi|^2$ in the planes $x=0$, $y=0$ and $z=0$.  The triangular Abrikosov flux tube array is visible on the $y=0$ plane, because the density drops to  $|\psi|^2\leq5n_n$ in the cores of the flux tubes relative to a background value of $|\psi|^2\geq8n_n$. Vortices are situated at $|\psi|^2\approx0$. In the $z=0$ and $x=0$ planes, we see low-$|\psi|^2$ troughs overlapping the flux tube array. Figure \ref{fig:backgrounddensity} confirms the existence of low-$|\psi|^2$ ($0\leq|\psi|^2 \leq 4n_n$) junctions shaped like crosses, where vortices and flux tubes intersect (e.g. $x=0$ slice), corresponding to the "spikes" visible in Figure \ref{fig:vortexcrystal}(b).

The locations of vortex filaments are identified precisely in Section \ref{section:tangle}. Key properties (vortex length $L$, average curvature $\langle \kappa \rangle$ and tangle polarity $\mathbf{p}$) are extracted. 

\section{Spin-Down-Driven Evolution}

\label{section:vortexevo}


 \begin{figure*}
\centering
\hspace*{-3em}
\includegraphics[height=23em]
{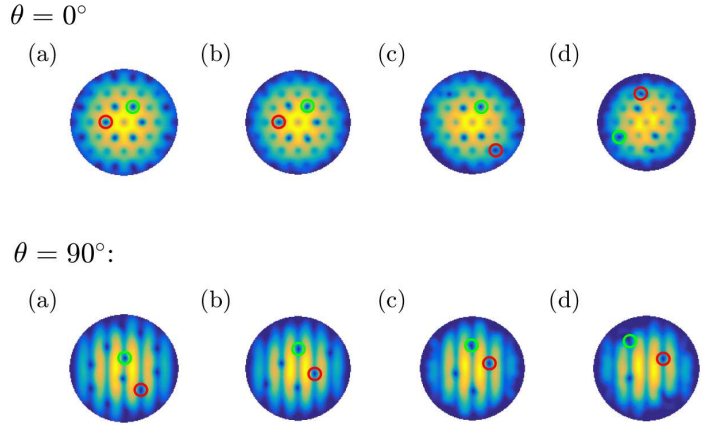}
\caption[Protons]
{Snapshots of neutron density in a cross-section through the midplane $z=0$, as the trap and co-rotating pinning grid spin down. \textit{Top}. $\textbf{B}$ and $\boldsymbol{\Omega}$ aligned ($\theta=0^{\circ}$). The triangular array of blue dots are the pinning sites. \textit{Bottom}. $\textbf{B}$ and $\boldsymbol{\Omega}$ misaligned ($\theta=90^{\circ}$). The blue striations are the pinning sites. The snapshots are taken at (a) $t=27\tau$, (b) $t=56\tau$, (c) $t=84\tau$ and (d) $t=111\tau$. Parameters: $\Omega(t=0)=0.5$, $\eta=-100$, $d_{\Phi}=2$, $\gamma=0.001$, $N_{\mathrm{em}}=-0.006I_c$, $\tilde{N}_n=N_n/(n_n\xi_n^{3})=8\times10^3$ and $\tilde{N}_p=0.05\tilde{N}_n$. The units are $n_n$ for all panels. Initially the system contains 18 vortices. The red and green circles follow the trajectories of two arbitrary vortices.}
\label{fig:slices}
\end{figure*}

 \begin{figure*}
\centering
\includegraphics[height=23em]
{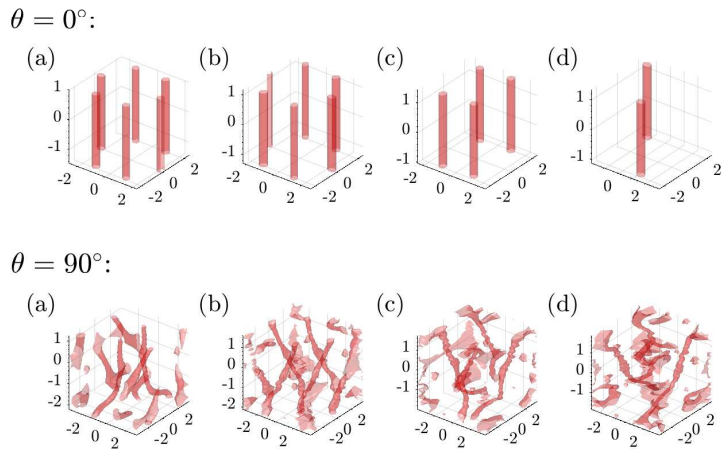}
\caption[Protons]
{Close-up snapshots of the vortex configurations in three dimensions for the system in Figure \ref{fig:slices}. Each panel displays the neutron density $|\psi|^2$ (in units of $n_n$). The red shading signifies the vortices, drawn where $|\psi|^2$ drops below 10\% of its maximum. Length units are $\xi_n$. We plot a close-up of the system, containing the six central vortices only, to avoid overcrowding the plot.}
\label{fig:isoevolution}
\end{figure*}

 \begin{figure}
\centering
\hspace*{-3em}
\includegraphics[height=15em]
{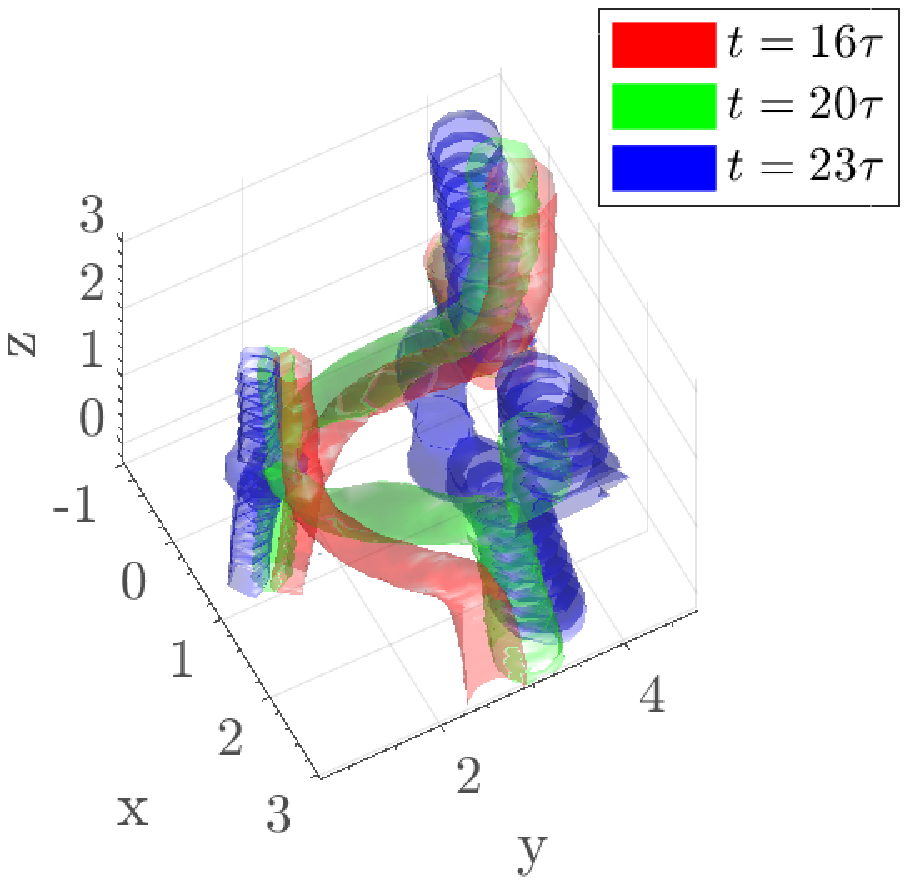}
\caption[Protons]
{Vortex reconnection event. Close-up snapshots of the neutron vortices for $\theta=90^{\circ}$. The snapshots are taken at $t=16\tau$ (red), $t=20\tau$ (green) and $t=23\tau$ (blue). The red, blue and green shading signifies the vortices, drawn where $|\psi|^2$ drops below 10\% of its maximum. Length units are $\xi_n$. Parameters: $\Omega(t=0)=0.5$, $\eta=-10$, $d_{\Phi}=3$, $\gamma=0.001$, $N_{\mathrm{em}}=-0.006I_c$, $\tilde{N}_n=N_n/(n_n\xi_n^{3})=8\times10^3$ and $\tilde{N}_p=0.05\tilde{N}_n$.}
\label{fig:reconnection}
\end{figure}

\subsection{Sporadic, step-wise unpinning}
As the stellar crust spins down due to the electromagnetic braking torque $N_{\mathrm{em}}$, the condensate loses angular momentum $\langle L_z \rangle$. In the absence of pinning, the vortices spiral outwards to the edge of the condensate and annihilate at the boundary to reduce $\langle L_z \rangle$. This behaviour can be seen in Figure \ref{fig:slices}. In the presence of pinning, the vortices unpin and repin one or more times in a step-wise fashion as they spiral outwards. In the top row of Figure \ref{fig:slices}, the triangular array of blue dots are the pinning sites, while in the bottom row of Figure \ref{fig:slices}, the blue striations are the pinning sites. To illustrate the step-wise motion, we present two examples for the convenience of the reader. The red and green circles follow the trajectories of two different vortices for both $\theta=0^{\circ}$ and $\theta=90^{\circ}$. For $\theta=0^{\circ}$, once the vortex circled in red unpins, it steps $|\Delta \phi| = 129^{\circ}$ azimuthally and $\Delta r = 2.28\xi_n$ radially between $t=56\tau$ and $t=84\tau$ and steps $|\Delta \phi| = 151^{\circ}$ azimuthally and $\Delta r = -0.697\xi$ radially between $t=84\tau$ and $t=111\tau$. It unpins and repins four times between $t=56\tau$ and $t=84\tau$ and five times between $t=84\tau$ and $t=111\tau$. For $\theta=90^{\circ}$, the part of the vortex sitting in the midplane ($z=0$) tends to slide along the $x=\mbox{constant}$ striations, where the flux tubes lie. For example, for $\theta=90^{\circ}$, the vortex circled in red remains pinned in the midplane for the duration of the simulation, with $\Delta x=0$ throughout. It steps by $\Delta y=1.31\xi$ between $t=56\tau$ and $t=84\tau$. The vortex circled in green does unpin and repin in the midplane, stepping by $\Delta x=2\xi$ and $\Delta y=0.503\xi$ between $t=84\tau$ and $t=111\tau$.

As in Section \ref{section:vortexcrystal}, we quantify the extent to which vortices are pinned to flux tubes by identifying contiguous subsets of points along the vortex. We then look to see how the fraction of pinned vortex points evolves. We classify a vortex point as pinned to a flux tube, if the vortex point lies within $0.4\xi_n$ of a flux tube point. Figure \ref{fig:pinnedfraction} plots the evolution of the fraction of vortex points that are pinned to the flux tube array out of the total initial vortex points. At $t=0$, we observe that all vortices are pinned for $\theta=0^{\circ}$, while only 0.210 of all vortex points are pinned for $\theta=90^{\circ}$. For $\theta=0^{\circ}$, the pinned fraction stays at 100\% until $t=18\tau$, whereupon it decays in a step-wise manner, as vortices unpin and either leave the condensate or remain as free vortices. The pinned fraction occasionally spikes upwards as vortices repin; for example at $t=26\tau$ it jumps from $0.778$ to $0.889$. By $t=60\tau$, the pinned fraction reaches $0.222$.  Fluctuations continue to occur until the end of the simulation due to vortices unpinning and repinning sporadically. For $\theta=90^{\circ}$, the initial pinned fraction is much smaller (0.210). As for $\theta=0^{\circ}$, the pinned fraction remains fairly constant until around $t=20\tau$, whereupon the fraction begins to decay gradually; it drops from $0.189$ to $0.0978$ during the period $20\tau\leq t\leq40\tau$. Repinning events break the monotonic decay. By $t=60\tau$, the fraction decays to $0.0408$.

\subsection{Vortex dynamics and reconnection}
The $\theta=90^{\circ}$ system does not preserve axial symmetry as it spins down. The vortices are tangled; the $z=\mbox{constant}$ cross-sections vary (see bottom row of Figure \ref{fig:isoevolution}), and vortices reorganise themselves following reconnections into complex tangles with nontrivial geometrical attributes. By contrast, the vortices remain rectilinear for $\theta=0^{\circ}$ (see top row of Figure \ref{fig:isoevolution}). For $\theta=0^{\circ}$, the vortices "hop" between the flux tubes in a radially outward direction while remaining straight. For $\theta=90^{\circ}$, different sections of a vortex experience different forces. One part of a vortex is held back due to stronger pinning forces at that location, whereas another part is freer to travel radially outwards. The vortex bends and becomes progressively more distorted, as the unpinned vortex segment moves further away from the pinned segment. Consequently, the flux tubes can "tear" the vortices. Due to Helmholtz's Second Theorem\footnote{A vortex cannot terminate in a fluid; it must either extend to the boundaries or form a closed loop.}, it is impossible for the vortex to truly separate into multiple segments at the pinning site (unless the density vanishes completely at that location). Instead the dynamics are restricted to the following options: (i) the unpinned segments drag the pinned segment away from the pinning site, and the vortex moves outwards in one piece; (ii) there is another vortex close enough, and the two vortices reconnect and exchange "tails"; (iii) the condensate boundary is close enough for the broken vortex segments to terminate there; or (iv) the unpinned and pinned segments break apart but each segment reforms a closed loop. 

In the simulations, phenomena (i), (ii) and (iii) are observed. Phenomenon (iv) is not observed; "vortex half-rings" intersecting the condensate boundary form frequently, but we see no complete vortex rings. Typically, a vortex close to the condensate boundary breaks at the flux tubes sites, and each broken segment forms a half-ring intersecting the condensate boundary. We observe "true" reconnections involving an exchange of tails between two different vortices as well as several instances of a vortex breaking at the site of the flux tubes and reconnecting with itself rather than exchanging tails with another vortex. Figure \ref{fig:reconnection} shows a close-up of the vortices before and after a true vortex reconnection event occurring at $t=22\tau$. 

The vortex half-rings that intersect the boundary in our simulations resemble the hairpin vortices occurring at turbulent boundary layers in classical fluids \citep{Hairpin}. In a similar fashion, \citet{Baggaley2013} observed vortex rings reconnecting with the walls of the container in the presence of a thermal counterflow in superfluid helium, and \citet{StaggParker2017} observed vortex loops partially attached to the boundary when modelling superfluid flow over a rough surface. 

\subsection{System size}
The system in Figures \ref{fig:slices} and \ref{fig:isoevolution} is small ($\leq18$ vortices). Different kinds of vortex-vortex interactions can occur in finite systems due to boundary-induced effects, such as double reconnections, rebounds and ejections \citep{Serafini2017}. Therefore, it is important to probe a larger system where boundary effects are less prominent. 

In a larger system, as in a neutron star, reconnections with other vortices or the formation of vortex loops should be more probable mechanisms than the interaction of vortices with the condensate boundary. Neighbouring vortices would are further apart in a real neutron star ($N_v/N_{\Phi}\sim10^{14}$), but vortex reconnections, which occur dynamically when vortices approach within a few core lengths of each other \citep{Koplik1993}, are still expected to occur. There is evidence for Vinen-type turbulence due to axial flow in our simulations [see Section \ref{section:tangle} and also \citet{DrummondMelatos2017}]. With increasing axial flow, the Donnelly-Glaberson instability develops, and Kelvin waves (helical perturbations of the vortex cores) grow exponentially \citep{Glaberson1974}. In Vinen-type turbulence, the total vortex line length obeys $L\propto v_{ns}^2$, where $v_{ns}$ is the counter-flow velocity. This result has been shown experimentally \citep{Swanson1983} and numerically \mbox{\citep{Tsubota2004}} for He-II. We expect that in a system exhibiting Vinen-type turbulence, the vortex crossing probability is high, due to the increase in vortex line density and subsequent generation of a dense vortex tangle. For example, \citet{Swanson1983} observed vortex line densities up to $L \approx 2500 \ \mbox{cm}^{-2}$ in He-II, cf. $10^4(\Omega/ \mbox{rad}\ \mbox{s}^{-1})^{-1}\mbox{cm}^{-2}$ in a typical neutron star. We will investigate in future work whether loop production through self-reconnecting vortices is more common, where the condensate is larger, and vortices are more widely separated. 

\citet{Bland2017} investigated a state that shares some similarities with our set-up. In their case, vortices in a dipolar BEC pin to density fluctuations, which act as pinning layers, rather than flux tubes. They observed large vortex loops simultaneously pinned to separate pinning layers. The vortex loops decay into two loops via reconnection, when the layers move apart. A similar phenomenon may occur at the flux tubes in a neutron star. The behaviour may change, once we allow the static flux tubes to move; the vortices may drag the flux tubes outwards rather than breaking up themselves. Further simulations need to be conducted to fully investigate this phenomenon.

 \section{Geometrical evolution of the vortex tangle} 
  \label{section:tangle}
The system we explore in this article is: (i) rotating, which promotes order, as vortices tend to crystallise in an array along $\boldsymbol{\Omega}$; (ii) coupled to a flux tube array misaligned with $\boldsymbol{\Omega}$, which introduces a superfluid flow along the rotation axis, which excites Kelvin waves \mbox{\citep{DrummondMelatos2017}} and (iii) coupled to a decelerating container, which means vortices are pushed through the static flux tube array as they spiral radially outwards. The combination of these three features produces a new kind of vortex tangle, namely a mixture of polarised quantum turbulence due to rotation and unpolarised grid turbulence due to the stresses experienced by the vortices as they are pushed against the rigid flux tube array, as suggested by \mbox{\citet{Link2012}}. Polarised quantum turbulence is the intermediate state between a polarised vortex array and isotropic turbulence \citep{JouMongiovi2008}. The Donnelly-Glaberson instability is a canonical example of polarised quantum turbulence. Axial flow exceeding a critical threshold in a rotating superfluid triggers the Donnelly-Glaberson instability  \mbox{\citep{Glaberson1974,donnelly1991}}. Grid turbulence occurs when a grid is steadily pushed through a superfluid, generating a slug of high-vorticity flow behind the towed grid \mbox{\citep{GridTurb,Barenghi2001,Krstulovic2016}}. 

 \begin{figure}
\centering
\vspace*{-1em}
\includegraphics[height=61em]
{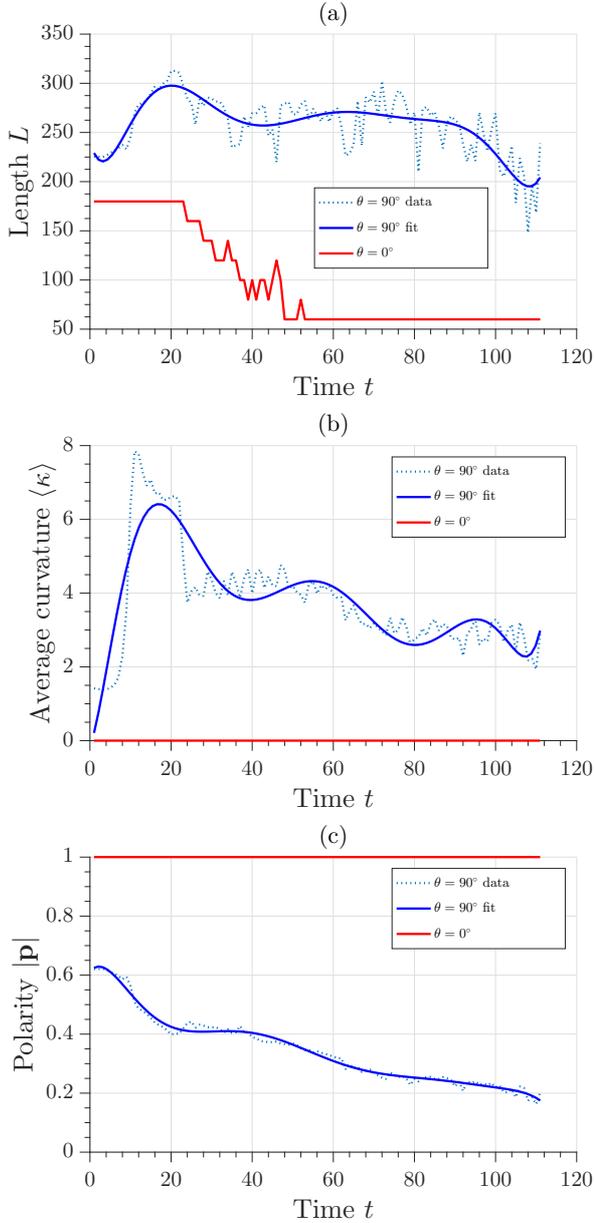}
\caption[Protons]
{Geometrical properties of the vortex tangle versus time, as the condensate spins down. (a) Total line length $L$ (in units of $\xi_n$). (b) Average curvature $\langle \kappa \rangle$ (in units of $\xi_n^{-1}$). (c) Polarity (dimensionless). The properties are graphed versus $t$ (in units of $\tau$) for $\theta=90^{\circ}$ (blue curves) and $\theta=0^{\circ}$ (red curve). The dotted blue curve indicates the data extracted using the vortex finding algorithm, while the solid blue curve gives a polynomial fit to the data. Parameters: $\Omega(t=0)=0.5$, $\eta=-100$, $d_{\Phi}=2$, $\gamma=0.001$, $N_{\mathrm{em}}=-0.006I_c$, $\tilde{N}_n=N_n/(n_n\xi_n^{3})=8\times10^3$ and $\tilde{N}_p=0.05\tilde{N}_n$.}
\label{fig:vortexproperties}
\end{figure}
 
A vortex finding algorithm is used to identify the point where a vortex intersects the plane $z=\mbox{constant}$ [\mbox{\citet{Jamesprivate}}; see also \mbox{\citet{Douglass2015}}]. We connect nearest neighbour intersection points in adjacent planes. A vortex terminates, where no neighbour exists within a given distance threshold. We identify contiguous filaments using a depth-first-search algorithm and use a smoothing spline function to approximate vortex cores as unbroken filaments. Certain geometrical quantities are used widely to characterise the degree of vortex tangling. We measure three: the total length,
\begin{equation} 
\label{eq:length}
L=\int \mbox{d} \xi \ |\mathbf{s'(\xi)}|,
\end{equation}
the length-averaged curvature,
\begin{equation}
 \label{eq:curvature}
\langle \kappa \rangle=\frac{1}{L}\int \mbox{d} \xi \ \frac{| \mathbf{s}'(\xi)\times \mathbf{s}''(\xi)|}{|\mathbf{s}'(\xi)|^2},
\end{equation}
and the length-averaged polarity,
\begin{equation} 
\label{eq:polarity}
\boldsymbol{p}=\frac{1}{L} \int \mbox{d} \xi \ \mathbf{s'(\xi)}.
\end{equation}
In (\ref{eq:length}), (\ref{eq:curvature}) and (\ref{eq:polarity}), $\mathbf{s}(\xi)$ denotes the displacement of an arbitrary point on the filament from the origin, labelled with the affine parameter $0\leq\xi\leq1$. A prime denotes differentiation with respect to $\xi$. 

We calculate $L$, $\langle \kappa \rangle$ and $\mathbf{p}$ as functions of time to understand how the vortex tangle evolves. The results are presented in Figure \ref{fig:vortexproperties}. The dotted blue curves in Figure \ref{fig:vortexproperties} indicates the data extracted using the vortex finding algorithm while the solid blue curve gives a polynomial fit to the data. 

\subsection{Length} 
Typically, $L$ increases when Kelvin waves are generated. When the Kelvin wave amplitude is large enough, neighbouring vortices reconnect. Figure \ref{fig:vortexproperties}(a) shows that $L$ is constant in the $\theta=0^{\circ}$ system, except when a vortex leaves the system\footnote{The spikes occurring at $t=34\tau$, $40\tau$, etc. correspond to vortices reentering the condensate after sitting just inside the condensate boundary (where they are undetectable by our vortex-finding algorithm) without fully exiting the system.}. We observe that vorticity begins to be extinguished at the boundary at $t\geq23\tau$ for $\theta=0^{\circ}$. By contrast, for $\theta=90^{\circ}$, $L$ is initially larger than for $\theta=0^{\circ}$, because the vortices zig and zag across the condensate as they pin to the flux tube lattice. The zigs and zags have a length-scale of order $d_{\Phi}$. The braking torque drives a further increase in $L$, as Kelvin waves are generated. The length decreases at $t\geq20\tau$, after vortices exit the system and annihilate at the boundary.  $L$ plateaus for $40\tau\lesssim t \lesssim90\tau$ and drops off after $t\geq90\tau$ at the end of the simulation. Significant fluctuations in $L$ over the simulation run $0\leq t\leq111\tau$ are caused by interactions with the boundary and reconnections.
 
Vortex length is best combined with other measures, such as $\langle \kappa \rangle$ and $|\mathbf{p}|$, to fully describe the tangle. For example, for $\theta=0^{\circ}$, $L$ decreases over time, as vortices leave the condensate, but $\langle \kappa \rangle=0$ and $|\mathbf{p}|=1$ remain constant, implying that there is no real change in the geometry.

\subsection{Curvature} 
The curvature $\langle \kappa \rangle$ measures how much the vortices crinkle; the average radius of curvature is $1/\langle \kappa \rangle$. Figure \ref{fig:vortexproperties}(b) shows how $\langle \kappa \rangle$ evolves, as the star spins down. The red curve corresponds to $\theta=0^{\circ}$, and the blue curve corresponds to $\theta=90^{\circ}$. For $\theta=0^{\circ}$, the vortices remain perfectly straight. For $\theta=90^{\circ}$, we find $\langle \kappa \rangle=0.21\xi_n^{-1}$ at $t=0$. This is roughly consistent with bending on the length-scale of the inter-flux-tube spacing $d_{\Phi}$. The flux tubes are spaced by $2\xi_n$; if the vortices snaked perfectly between every flux tube this would yield  $\langle \kappa \rangle=0.5\xi^{-1}$. The actual measured $\langle \kappa \rangle$ is smaller than this, because the vortices are partially polarised by the rotation. Kelvin wave generation causes $\langle \kappa \rangle$ to increase, until it peaks at $t=11\tau$, reaching $\langle \kappa \rangle=7.85\xi_n^{-1}$. $\langle \kappa \rangle$ drops off steeply for $t\geq11\tau$ and continues to decay for the remainder of the simulation run $11\tau\leq t \leq111\tau$.  Fluctuations in $\langle \kappa \rangle$ occur due to vortex-vortex interactions (including reconnections) and vortex-boundary interactions.

\subsection{Polarity} 
We can decompose the tangle into polarised and unpolarised (isotropic) components. A non-zero polarity $\mathbf{p}$ indicates polarisation, because perfectly random orientations would cancel when summed. A perfectly polarised tangle satisfies $|\mathbf{p}|=1$, because all tangents $\mathbf{s'(\xi)}$ lie parallel to each other. In Figure \ref{fig:vortexproperties}(c), we find $|\mathbf{p}|=1$ throughout the simulation for $\theta=0^{\circ}$, as the vortex array remains rectilinear. For $\theta=90^{\circ}$, however, $|\mathbf{p}|$ decreases from $0.625$ to $0.175$; the tangle becomes more isotropic, as the condensate spins down. If we ran a longer simulation, it is possible that $|\mathbf{p}|$ would asymptote to a partially polarised state with $|\mathbf{p}|>0$. Alternatively, $|\mathbf{p}|$ could continue to decay until the system is entirely isotropic. The former seems more probable: the rotation should keep the system at least partially polarised for the duration of the simulation \citep{JouMongiovi2008}.

The time-averaged angle between $\mathbf{p}$ and $\boldsymbol{\Omega}$ is $\langle\mbox{cos}^{-1}(\boldsymbol{\Omega}\cdot\mathbf{p}/|\boldsymbol{\Omega}||\mathbf{p}|)\rangle=0.0314^{\circ}$,  where $\langle\ldots\rangle$ here denotes an average over the simulation run $0\leq t\leq111\tau$. Even though $|\mathbf{p}|$ decays over the course of the simulation (i.e. less of the vorticity is aligned along $\boldsymbol{\Omega}$), the angle between $\mathbf{p}$ and $\boldsymbol{\Omega}$ does not vary significantly, and $\mathbf{p}$ remains closely aligned with $\boldsymbol{\Omega}$ throughout. The vector components of $\mathbf{s}'(\xi)$ perpendicular to $\boldsymbol{\Omega}$ mostly cancel.

\mbox{\citet{Tsubota2004}} studied vortex tangles in He-II. Their initial configuration is a polarised array ($|\mathbf{p}|=1$) which becomes tangled due to a counterflow. We observe that their set-up is different to ours: their trap does not decelerate. Nonetheless, their results resemble ours in a few repects. For  $\Omega=4.98\times10^{-3}\ \mbox{rad}\ \mbox{s}^{-1}$, $|\mathbf{p}|$ decays from 1 to 0.4 after $100\ \mbox{s}$, while $L$ increases from $35\ \mbox{cm}^{-2}$ to $90\ \mbox{cm}^{-2}$ after $70 \ \mbox{s}$ and subsequently decays to $72 \ \mbox{cm}^{-2}$ after $350\ \mbox{s}$.

To summarise, for $\theta\neq0^{\circ}$, the braking torque initially drives Kelvin wave generation, and the crinkling and isotropy of the vortex tangle increase. When vortices begin to annihilate at the condensate boundary (at $t\geq20\tau$), $L$, $\langle \kappa \rangle$ and $|\mathbf{p}|$ decrease. We do not observe the emergence of a random unstructured tangle; rather, the tangle remains polarised by the rotation, decreasing from $|\mathbf{p}|=0.625$ initially to $0.175$ at $t=111\tau$, corresponding to an increasingly isotropic distribution of vorticity. 

 \section{Feedback on the Crust} 
 \label{section:rotation}




There are long-standing claims in the neutron star literature that neutron vortices and proton flux tubes are strongly pinned for $\theta=0^{\circ}$, affecting the crust's ability to spin down \citep{Srivasan1990,Ruderman1998,KonarBhatta1999,KonenkovGeppert2001}. So it is natural to ask: by how much does pinning reduce the observable $\dot{\Omega}$ as a function of $\eta$ and $\theta$? Is the vorticity transported to the condensate boundary more "easily" in a vortex tangle ($\theta\neq0^{\circ})$, because the vortices can slip around the flux tubes? Or is vorticity transport "harder", because vortices slip past each other without necessarily triggering nearest neighbour unpinning avalanches by proximity knock-on \citep{LilaMelatos2011,LilaMelatos2013,HaskellMelatos2016}? 

\begin{figure}
\centering
\includegraphics[height=40em]
{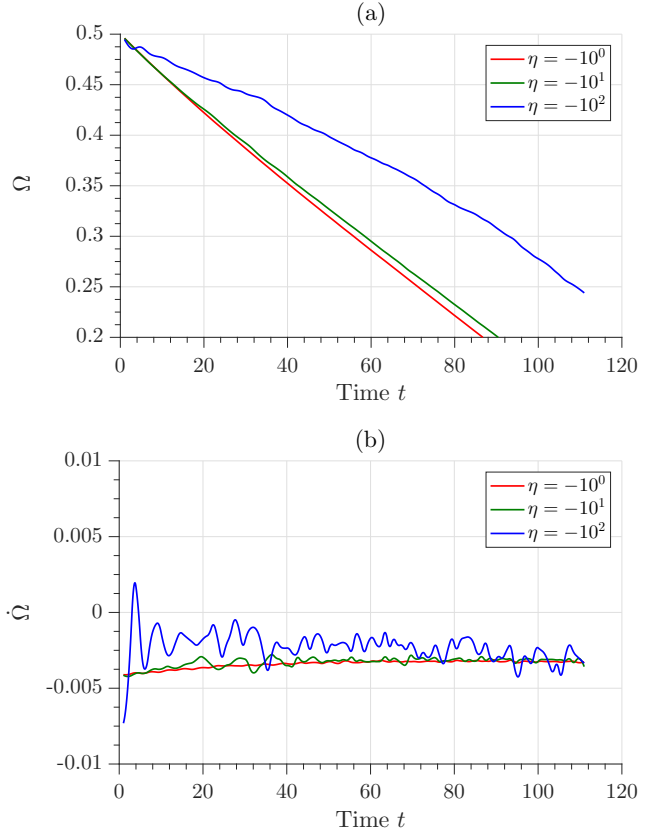}
\caption
{Feedback on the crust versus pinning potential. Crust angular velocity $\Omega(t)$ (\textit{top}) and rate of change of crust angular velocity $\dot{\Omega}(t)$ (\textit{bottom}) as functions of time (in units of $\tau$) for $\eta=-10^0$ (red curve), $\eta=-10^1$ (green curve) and $\eta=-10^2$ (blue curve). Parameters: $\Omega(t=0)=0.5$, $\theta=60^{\circ}$, $d_{\Phi}=2$, $\gamma=0.001$, $N_{\mathrm{em}}=-0.006I_c$, $\tilde{N}_n=N_n/(n_n\xi_n^{3})=8\times10^3$ and $\tilde{N}_p=0.05\tilde{N}_n$.}
\label{fig:omegapinning}
\end{figure}

In Figure \ref{fig:omegapinning}, the crust's spin-down rate including feedback decreases, as $|\eta|$ increases. We quantify this by computing the time-averaged rate of change of the angular velocity $\langle\dot{\Omega}\rangle$. Here, $\langle\ldots\rangle$ denotes an average over the whole simulation run $0\leq t\leq111\tau$. Pinning impedes the deceleration of the crust: the average spin-down rate $\langle \dot{\Omega}\rangle$ is $-0.0034$ for $\eta=-1$, $-0.0033$ for $\eta=-10$ and $-0.0023$ for $\eta=-10^2$. When vortices pin strongly to the flux tubes (which co-rotate with the crust), the electromagnetic torque acts simultaneously on both the crust and condensate. When the pinning is weak, the electromagnetic torque spins down the crust unencumbered by pinning to the condensate. In the hypothetical case where there pinning between the crust and the superfluid vanishes, there would be no feedback from the superfluid and the crust would spin down at the precise rate dictated by the electromagnetic torque acting on it.

In addition, the variance in $\dot{\Omega}$ over the simulation run $0\leq t\leq111\tau$, denoted by $\sigma^2_{\dot{\Omega}}$, increases as $|\eta|$ increases [see Figure \ref{fig:omegapinning}(b)]. The standard deviations for $|\eta|=1$, $|\eta|=10$ and $|\eta|=10^2$ are $\sigma_{\dot{\Omega}}(\eta=1)=2.5\times10^{-4}$, $\sigma_{\dot{\Omega}}(\eta=10)=2.9\times10^{-4}$ and $\sigma_{\dot{\Omega}}(\eta=10^2)=9.1\times10^{-4}$. As the vortices sporadically unpin and repin, they move in and out of different pinning potentials. The deeper potentials exert larger forces, so variations in $\dot{\Omega}$ are larger for greater $|\eta|$.

\begin{figure}
\centering
\includegraphics[height=40em]
{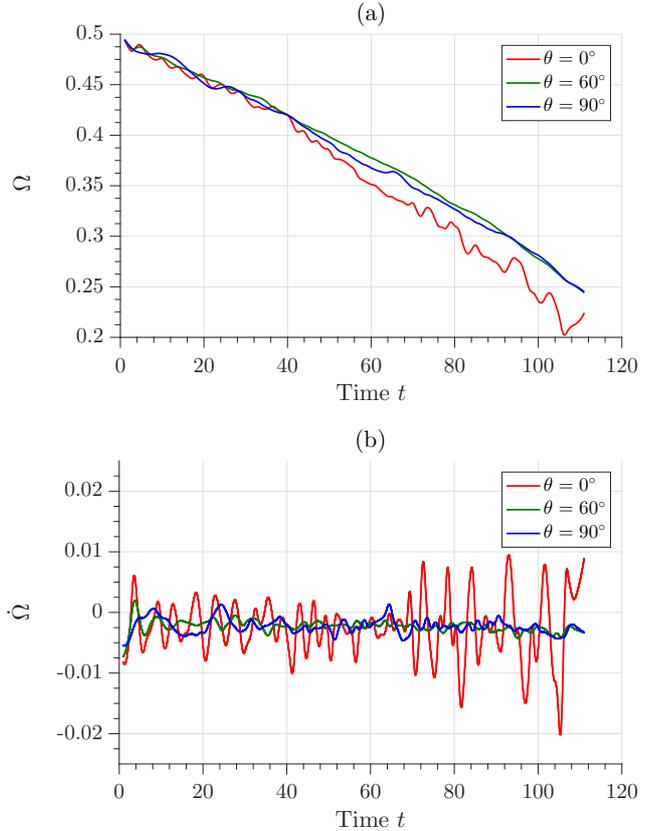}
\caption
{Feedback on the crust versus angle $\theta$ between the rotation and magnetic axes. Crust angular velocity $\Omega(t)$ (\textit{top}) and rate of change of crust angular velocity $\dot{\Omega}(t)$ (\textit{bottom}) as functions of time (in units of $\tau$) for $\theta=0^{\circ}$ (red curve), $\theta=60^{\circ}$ (green curve) and $\theta=90^{\circ}$ (blue curve). Parameters: $\Omega(t=0)=0.5$, $\eta=-10^2$, $d_{\Phi}=2$, $\gamma=0.001$, $N_{\mathrm{em}}=-0.006I_c$, $\tilde{N}_n=N_n/(n_n\xi_n^{3})=8\times10^3$ and $\tilde{N}_p=0.05\tilde{N}_n$.}
\label{fig:rotation}
\end{figure}

In Figure \ref{fig:rotation}, we observe that $\langle\dot{\Omega}\rangle$ behaves differently for $\theta=0^{\circ}$ and $\theta\neq0^{\circ}$. For $t\geq40\tau$, the $\theta=0^{\circ}$ and $\theta\neq0^{\circ}$ curves diverge. The percentage difference between the mean rates for $\theta=0^{\circ}$ and $\theta=90^{\circ}$ is $(\langle{\dot{\Omega}}\rangle_{\theta=0^{\circ}}-\langle{\dot{\Omega}}\rangle_{\theta=90^{\circ}})/\langle{\dot{\Omega}}\rangle_{\theta=90^{\circ}}=8.2\%$. This suggests that it is "harder" for the crust to spin down in the $\theta\neq0$ case; the vortices do slip around the flux tubes more easily, with $\langle L_z(t) \rangle$ decreasing more rapidly for $\theta\neq0$ than $\theta=0$. Note that we cannot rule out the possibility that this effect is an artefact of the small number of vortices left at this stage in the simulation. On the other hand and by contrast, the percentage difference between the mean rates for $\theta=60^{\circ}$ and $\theta=90^{\circ}$ is small: $(\langle{\dot{\Omega}}\rangle_{\theta=60^{\circ}}-\langle{\dot{\Omega}}\rangle_{\theta=90^{\circ}})/\langle{\dot{\Omega}}\rangle_{\theta=90^{\circ}}=0.41\%$, suggesting that tangled vorticity plays a key role.

The variance $\sigma^2_{\dot{\Omega}}$ is significantly greater for $\theta=0^{\circ}$ than $\theta\neq0^{\circ}$ [see Figure \ref{fig:rotation}(b)] with $\sigma_{\dot{\Omega}}(\theta=90^{\circ})=4.5\times10^{-3}$, $\sigma_{\dot{\Omega}}(\theta=60^{\circ})=9.1\times10^{-4}$ and $\sigma_{\dot{\Omega}}(\theta=90^{\circ})=1.2\times10^{-3}$ computed over the whole simulation run $0\leq t\leq111\tau$. In particular, changes in $\Omega(t)$ are spasmodic for $\theta=0^{\circ}$, while $\Omega(t)$ varies more smoothly for $\theta\neq0^{\circ}$. For $\theta=0^{\circ}$, the vortex motion is restricted: a vortex either pins to a flux tube or stays unpinned, without any options in-between. A greater crust-superfluid lag must accumulate to unpin a vortex along its entire length, potentially knocking-on other vortices in the process. Hence, $\Omega(t)$ only changes in discrete bursts, corresponding to a vortex unpinning and then repinning at a new site. In contrast, for $\theta\neq0^{\circ}$, $\Omega(t)$ responds immediately and smoothly to $N_{\textrm{em}}$, as vortices unzip gradually. In other words, vortices move more freely when tangled. The vortices do not always unpin along their entire length, and some vortices are pinned simultaneously to multiple flux tubes [see Figure \ref{fig:pinnedfraction}(c)]. This result has interesting implications for the glitch phenomenon, which will be explored in future work.

\section{Limitations of the Model}
    \label{section:limitations}
In this section, we briefly discuss the limitations of the model foreshadowed in Section \ref{section:GPmodel}. We refer the reader to Section 5 of \citet{DrummondMelatos2017} for full details. The model involves four categories of idealisations: (i) we approximate the neutron superfluid in the outer core as a Bose-Einstein condensate, whose evolution is described by the GPE; (ii) we employ a static, prescribed ansatz for the proton order parameter; (iii) feedback on the crust is assumed to be instantaneous; and (iv) certain parameter values in the simulations are astrophysically unrealistic. Therefore one must proceed with caution when applying the results to astrophysical data.

(i) \textit{GPE approximation.} The GPE model has been employed successfully in previous work to simulate the microphysics of neutron star glitches, producing avalanche statistics consistent with observational data \citep{LilaMelatos2011,Douglass2015}. Yet the GPE describes a dilute, weakly interacting Bose gas, whereas neutron star matter is fermionic, strongly interacting and not dilute.  The Bogoliubov De Gennes (BdG) equations, which describe condensed Fermi gases, are also restricted to the weakly-interacting regime and are too expensive to solve numerically given the resources at our disposal \citep{Han2010}. In the cross-over region, there is a smooth transition from the GPE to the BdG equations \citep{Zwerger2016}. Here the scattering length and average neutron-neutron spacing are of roughly the same order, more closely approximating the regime inside a neutron star \citep{Chang2004}.  

The trapping potential $V$ serves the important function of preventing the rotating BEC from escaping the simulation box. Previous GPE simulations demonstrate that the exact form of the trap (hard-wall or harmonic) does not affect the microphysics of vortex unpinning \citep{LilaMelatos2011, LilaMelatos2012, Douglass2015}.

(ii) \textit{Proton ansatz.} The proton response to neutron motion is not included in our model. Hence entrainment of the protons around the neutron vortices does not occur, and the magnetic field associated with the neutron vortices does not contribute to $\mathcal{H}_{int}$. In addition, the proton flux tubes are rigid and fixed in this paper, whereas in reality  they are likely to become distorted in response to spin down, perhaps even forming a tangle themselves. We hope to include the proton response in future work by solving Ampere's law along with the proton and neutron equations of motion. Unfortunately, this will increase further the computational load.

(iii) \textit{Feedback on the crust.}  Equation (\ref{eq:brake}) assumes changes in $\langle L_z(t) \rangle$ instantly affect the angular velocity of the crust. In reality, these changes are communicated with a time lag by sound waves and Kelvin waves \citep{Jones1998}. Although these processes are not instantaneous, they are nonetheless fast \citep{LilaMelatos2011}. 

(iv) \textit{Parameter values.} The simulated parameter ranges are very different to the relevant astrophysical values. For example, the size of the simulation box is $10^{-12}\ \mbox{m}$, while the diameter of a canonical neutron star is $20 \ \mbox{km}$. The number of vortices in the simulations is $N_v\approx 20$, cf. $N_v=10^{16}(\Omega/ \mbox{rad} \ \mbox{s}^{-1})$ in a neutron star. \citet{DrummondMelatos2017} estimated the dimensionless density coefficient to be $|\eta|\approx0.2$ in the outer core. In this paper, we investigate $\eta$ values in the range $1<|\eta|<10^2$. However, it should be noted that the theoretical estimate of $|\eta|$ is approximate and considerable work needs to be done to refine it with reference to neutron star conditions. It is important to note that the realistic parameter regime for a neutron star will remain inaccessible computationally for many years. We are forced to use a much smaller dynamic range to make a start on this challenging problem.

\section{Conclusions}
The chief aim of this paper is to characterise the far-from-equilibrium vortex behaviour in a pinned, decelerating condensate, when $\mathbf{\Omega}$ and $\boldsymbol{B}$ are misaligned. In Section \ref{section:vortexcrystal}, we find that the initial vortex configuration for $\theta\neq0^{\circ}$ is a complex, partially polarised structure ("vortex crystal"), arising as a compromise between global vortex alignment along $\mathbf{\Omega}$ and local pinning along $\boldsymbol{B}$. In Section \ref{section:vortexevo}, we observe vortices unpinning and repinning sporadically in a step-wise fashion, as the container decelerates. For $\theta=0^{\circ}$, vortices hop between pinning sites as they spiral towards the condensate edge, while, for $\theta=90^{\circ}$, pinned vortex segments tend to slide along the flux tubes as they move outwards. For $\theta\neq0^{\circ}$, vortices are either (i) dragged in their entirety from the flux tube pinning sites, or (ii) are torn by the flux tubes and reconnect with other vortices or the container wall.

In Section \ref{section:tangle}, we characterise the polarised vortex tangle geometrically, as the container decelerates. The superfluid experiences grid turbulence as the vortices are pushed through the flux tube array, while the rotation of the superfluid polarises the tangle, with vortices tending to crystalise along the rotation axis.  We find that spin-down-driven tangling competes against vorticity loss at the boundary. The braking torque initially drives Kelvin wave generation and $L$ increases. Later on ($t\geq20\tau$), vortices annihilate at the condensate boundary and $L$ subsequently decreases; the tangle decays as the system loses vorticity. The fraction of vortex length pinned to the flux tubes decays throughout the simulation for both $\theta=0$ and $\theta\neq0$. 

Another key aim of the paper is to study whether vortices and flux tubes are locked together during spin down or slip past each other. We are not in a position to answer this question categorically, and the situation is likely to remain uncertain for some time given the computational challenge. Nevertheless, in Section \ref{section:rotation}, we find that increasing $|\eta|$ impedes the spin down of the crust. We observe smoother vorticity transport for $\theta\neq0^{\circ}$, where tangled vortices slip more easily around the flux tubes, in contrast to unpinning and re-pinning in discrete units for $\theta=0^{\circ}$. 

The occurrence of a glitch decreases the crust-superfluid lag and reduces the global unpinning rate temporarily. In this manner, the  system self-regulates to accommodate a persistent electromagnetic spin-down torque. This type of slowly driven, self-regulating system exhibits self-organised criticality \citep{LilaMelatos2013,Fulgenzi2017}. Due to computational limitations, the three-dimensional simulations in this paper contain too few vortices to observe truly collective behaviour such as the vortex avalanches seen in previous two-dimensional simulations \citep{LilaMelatos2011,LilaMelatos2012,LilaMelatos2013,Douglass2015}. Simulating a larger box containing $\geq10^2$ vortices would allow us to explore the statistical implications of vortex-flux-tube interactions for glitch sizes and waiting times. It is an open question as to whether the extra degree of freedom in three dimensions suppresses glitch activity or not. Knock-on processes are required to correlate the unpinning probability of a vortex with the pinning state of its neighbours \citep{LilaMelatos2013}. In future we hope to study how knock-on works, when the rotation and magnetic axes are misaligned ($\theta\neq0$). Finally, and most importantly, the flux tube response needs to be investigated, by solving for the proton and neutron dynamics simultaneously. For example, what role does flux tube motion play in determining glitch sizes, waiting times and $\langle \dot{\Omega}\rangle$?


\section*{Acknowledgements}
Simulations were conducted using the Multi-modal Australian ScienceS Imaging and Visualisation Environment (MASSIVE)\footnote{\url{https://www.massive.org.au}} at Monash University with access provided by a National Computational Infrastructure Merit Allocation Scheme grant. The project was supported by the Australian Research Council through a Discovery Project grant and the Centre of Excellence for Gravitational Wave Discovery (OzGrav; CE170100004).



\bibliographystyle{mnras}
\bibliography{MNRAS_fluxtubes_II} 


\bsp	
\label{lastpage}
\end{document}